\documentclass[aps, prb, twocolumn, amssymb, amsmath, showpacs, superscriptaddress]{revtex4}

\usepackage[T1]{fontenc}
\usepackage[latin9]{inputenc}
\usepackage{amsmath}
\usepackage{graphicx}
\usepackage{amssymb}
\usepackage{esint}
\usepackage{SIunits}
\usepackage[colorlinks,
            linkcolor=red,
            anchorcolor=green,
            citecolor=blue
            ]{hyperref}
\begin{document}

\title{Valley-polarized quantum anomalous Hall phases and tunable topological phase transitions in half-hydrogenated Bi honeycomb monolayers}

\author{Cheng-Cheng Liu}
\affiliation {School of Physics, Beijing Institute of Technology, Beijing 100081, China}

\author{Jin-Jian Zhou}
\affiliation {Institute of Physics, Chinese Academy of Sciences, Beijing 100190, China}
\affiliation {School of Physics, Beijing Institute of Technology, Beijing 100081, China}

\author{Yugui Yao}
\email{ygyao@bit.edu.cn}
\affiliation {School of Physics, Beijing Institute of Technology, Beijing 100081, China}

\begin{abstract}
Based on first-principles calculations, we find novel valley-polarized quantum anomalous Hall (VP-QAH) phases with a large gap-0.19 eV at an appropriate buckled angle and tunable topological phase transitions driven by the spontaneous magnetization within a half-hydrogenated Bi honeycomb monolayer. Depending on the magnetization orientation, four different phases can emerge, i.e., two VP-QAH phases, ferromagnetic insulating and metallic states.
When the magnetization is reversed from the +$\mathbf{z}$ to -$\mathbf{z}$ directions, accompanying with a sign change in the Chern number (from -1 to +1), the chiral edge state is moved from valley $K$ to $K'$. Our findings provide a platform for designing dissipationless electronics and valleytronics in a more robust manner through the tuning of the magnetization orientation.
\end{abstract}


\maketitle

\section{INTRODUCTION}

Recently, there has been broad interest in the condensed matter physics community in the search for novel topological phases, as well as the tuning and understanding of the related phase transitions, aiming for both scientific explorations and potential applications~\cite{Hasan2010,Qi2011}.
Among these novel topological phases is the quantum anomalous Hall (QAH) phase~\cite{Haldane1988,Onoda2003,Liu2008,Yu2010,Qiao2010}, which is characterized by a finite Chern number and chiral edge states in the bulk band gap, and maintains robust stability against disorder and other perturbations~\cite{Thouless1982}. Although the first proposal appeared over twenty years ago, 
not until recently was the experimental evidence for QAH phase reported within Cr-doped (Bi,Sb)$_2$Te$_3$ at extremely low temperatures~\cite{Chang2013,Kou2014}.
Some strategies to achieve QAH effects at high temperatures in honeycomb materials are proposed~\cite{Ding2011,Wang2013,Zhang2013,Garrity2013,Qiao2014,Zhang2012,Ezawa2012,Pan2014,Huang2014,Wu2014,Zhang2011}, such as by transition-metal atoms adaption~\cite{Qiao2010,Wang2013,Zhang2013}, magnetic substrate proximity effect~\cite{Garrity2013,Qiao2014}, and surface functionalization~\cite{Huang2014,Wu2014}.
In honeycomb lattices, $K$ and $K'$ valleys similar to real spin, provide another tunable binary degree of freedom to design valleytronics. By breaking the inversion symmetry, a bulk band gap can be opened to host a quantum valley Hall (QVH) effect, which is classified by a valley Chern number $C_v=C_K-C_{K'}$~\cite{Xiao2007,Martin2008,Gorbachev2014}.
For technological applications, it is important
to find a novel topological state with a large gap that simultaneously shares the properties of both QAH states and QVH states, i.e. valley-polarized QAH (VP-QAH) phases, on the one hand, and learn how to drive transitions among different phases, on the other.

The element bismuth has the largest SOC strength in the periodic table of elements except radioactive elements. The above exotic topological quantum phases can be expected to emerge notably in the Bi-based materials. Actually, the star materials for three-dimensional (3D) TI are no other than the Bi-based materials--Bi$_2$Se$_3$, and Bi$_2$Te$_3$~\cite{Zhang2009}. Bi-based material Cu$_x$Bi$_2$Se$_3$ is predicted as a 3D time-reversal-invariant topological superconductor~\cite{Fu2010}. The honeycomb monolayer Bi film is a two-dimensional (2D) TI with the SOC gap opened at $\Gamma$ point~\cite{Murakami2006,Liu2011}.  A fully-hydrogenated (F-H) or halogenated Bi honeycomb monolayer is predicted a 2D TI with a record bulk band gap (> 1 eV) at $K$ and $K'$ points~\cite{Song2014,Liu2014}. For electron doping, the 2D TI F-H monolayer is predicted to host time-reversal-invariant $p\pm ip$ topological superconductivity~\cite{Yang2014}.

\begin{figure}
\includegraphics[width=3.5in]{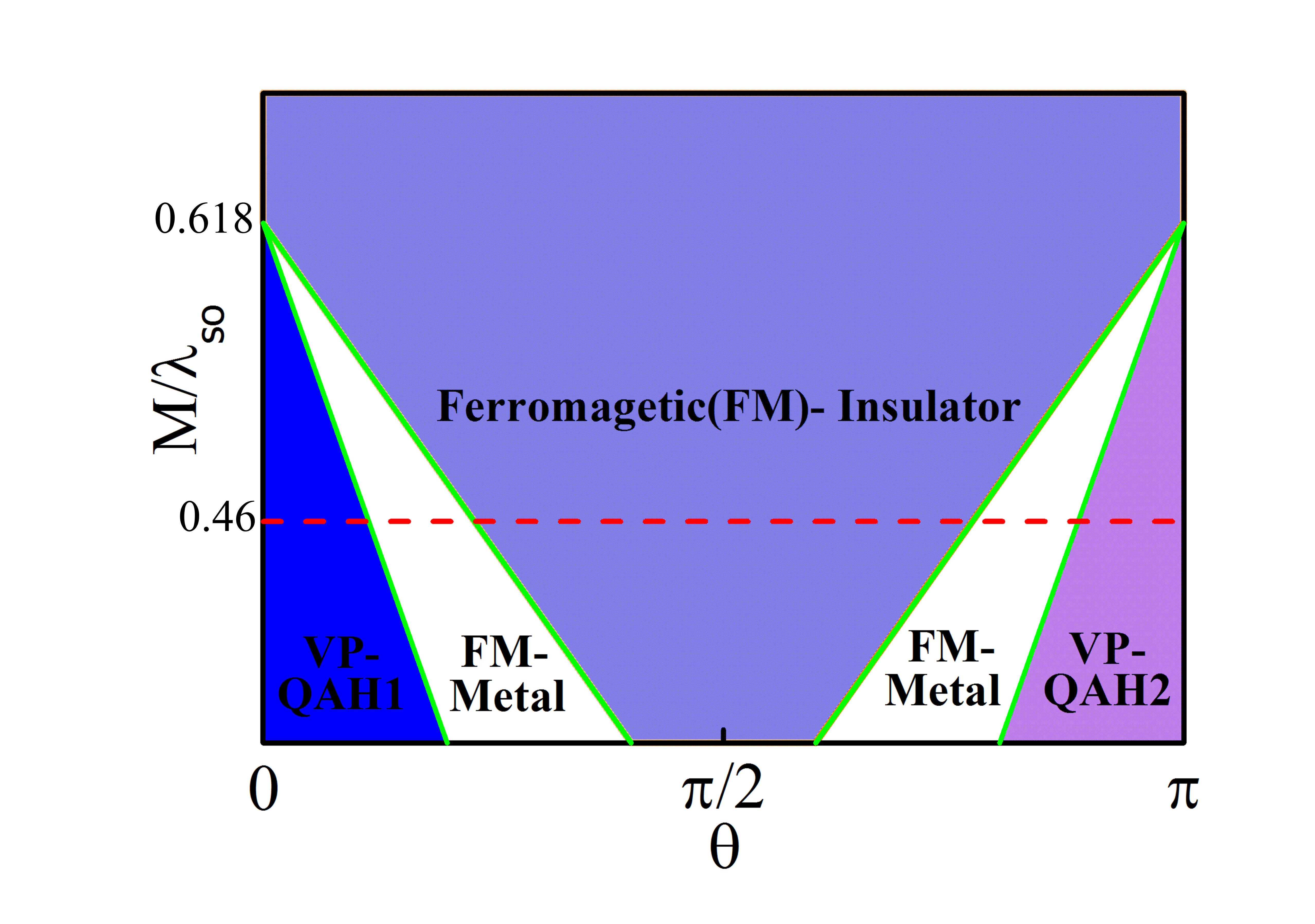}
\caption{(color online). A sketch of the predicted phase diagram as a function of the direction and magnitude of the magnetization in the H-H Bi honeycomb monolayer. $\theta$  and $M$  are the polar angles and magnitude of the spontaneous magnetization $\mathbf{M}$. $\lambda_{so}$ is half the intrinsic SOC strength.  VP-QAH1(2) corresponds to $C=-1$($C=1$) and $C_v=-1$($C_v=-1$). The two cases can be related by time reversal operation. The red dash line is the phase transition path occurred in the H-H Bi monolayer. }\label{fig:phase_diagram}
\end{figure}

Here, we report the theoretical finding of two novel VP-QAH topological phases, where QVH and QAH effects coexist, and associated topological phase transitions caused by the magnetization orientation in a half-hydrogenated (H-H) Bi honeycomb monolayer. The band gap of the VP-QAH phases can reach 0.19 eV with an appropriate buckled angle. Other phases, shown in Fig.~\ref{fig:phase_diagram}, such as ferromagnetic (FM)-Metal and FM-Insulator phases, are found in several regions with different magnetization orientations. The magnetization orientation can be tuned via an external magnetic field or proximity induction by different magnetic substrates. Furthermore, these common experimental measures can be used to control the topological phase transitions in such a H-H Bi monolayer. Therefore, our findings provide an ideal platform for the design of dissipationless electronics and valleytronics in a robust and controllable manner.

First-principles (FP) calculations are performed using the  projector augmented wave method implemented in the Vienna $\textit{ab initio}$ simulation package (VASP)~\cite{Kresse1996}.  Perdew-Burke-Ernzerhof parametrization of the generalized gradient approximation (GGA-PBE) is used for the exchange correlation potential~\cite{Perdew1996}. The plane wave energy cutoff is set to 300 eV, and the Brillouin zone is sampled by a $24\times24\times1$ mesh. The single layer structures were constructed with a vacuum layer of 20  \AA\   to avoid the interactions between the layers. By using the Wannier90 code, the maximally localized Wannier functions are constructed and the Berry curvature is obtained~\cite{Mostofi2008,Marzari1997,Souza2001}.  Based on the constructed Wannier functions, we use an iterative method~\cite{Sancho1985} to obtain the surface Green's function of the semi-infinite system, from which we can calculate the dispersion of the edge states.

\section{RESULTS AND DISCUSSION}

\begin{figure}
\includegraphics[width=3.5in]{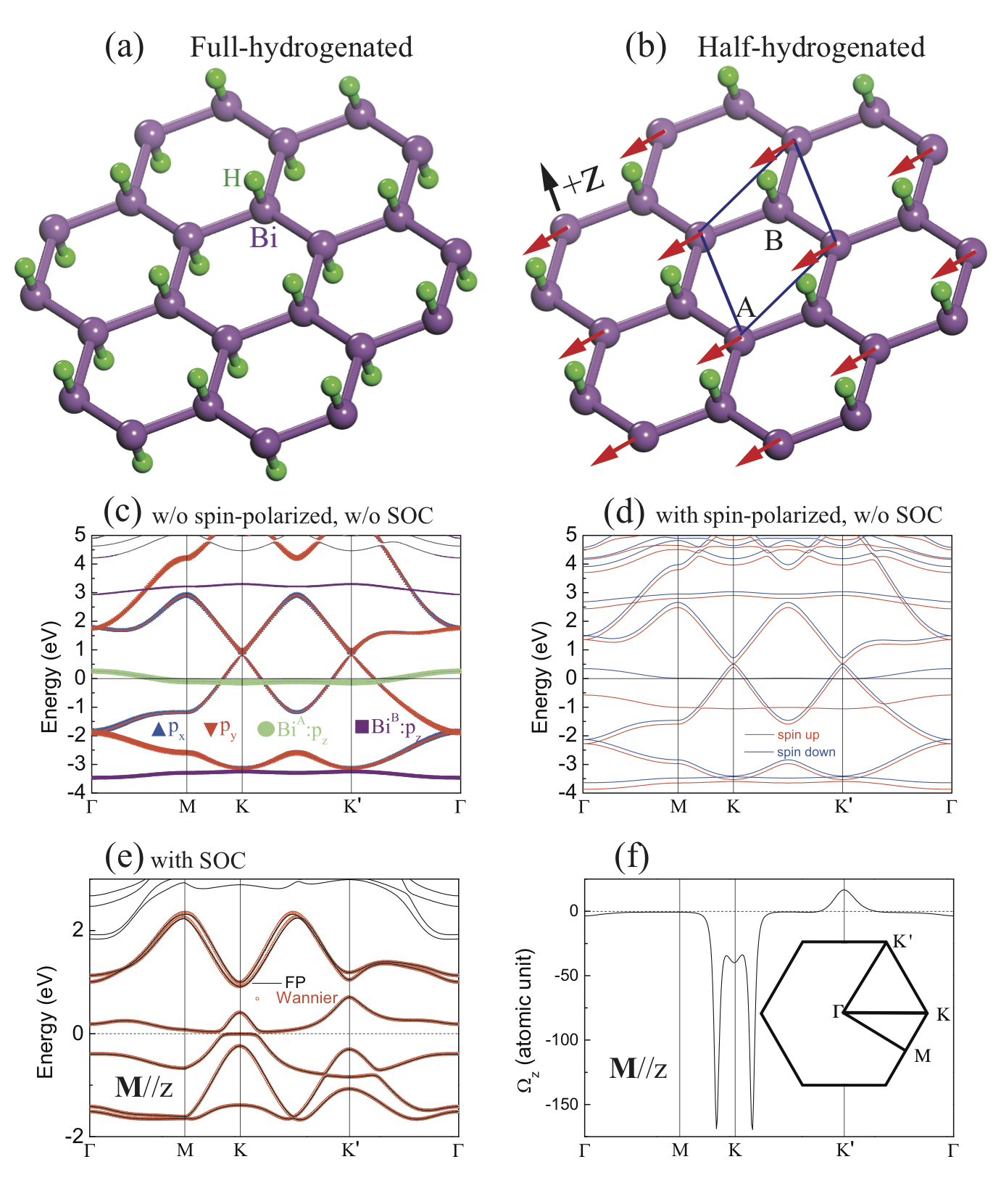}
\caption{(color online). (a),(b) The respective lattice geometries for the F-H and H-H Bi honeycomb monolayers. Large purple and small green spheres represent the Bi and hydrogen atoms, respectively. A and B in (b) label the two sublattices. The red arrows in (b) represent the magnetic moments. The black arrow marks the +$\mathbf{z}$ direction. (c) The projection band structures for the H-H Bi monolayer. The color of the symbols labels the different atomic orbitals, and their size is proportional to the weight of the band eigenfunctions on these atomic orbitals. (d) The spin-polarized band structure for the H-H Bi monolayer. (e) The SOC band structure for the magnetization along $\mathbf{z}$ axis. The red circle (black line) is for the bands from the Wannier interpolation (FP). Both are in good agreement. (f) The Berry curvature distribution along the line with high symmetry for the summation of all the valence bands. Inset: The first Brillouin zone and its points of high symmetry.}\label{fig:structure}
\end{figure}

Figure~\ref{fig:structure}(a) plots the typical geometries for a F-H Bi monolayer, with three-fold rotation symmetry and inversion symmetry, like for graphane. The H-H Bi monolayer with a quasi-planar geometry can be obtained by removing half of the hydrogenation of the F-H Bi monolayer~\cite{Song2014,Liu2014}. The lattice constant is 5.54 \AA\   with the distance between Bi and H atoms $d_{Bi-H}$=1.83 \AA .
Its three-fold rotation symmetry remains but inversion symmetry broken, as shown in Fig.~\ref{fig:structure}(b), similar to graphone. Figures~\ref{fig:structure}(c-e) plot the band structures for three cases within the GGA, spin-polarized and SOC calculations. As shown in Fig.~\ref{fig:structure}(c), there is a flat band near the Fermi level from the $p_z$ orbital of the Bi atoms disconnected from the H atoms, since these Bi atoms constitute a triangular lattice with large bond length. For spin-polarized calculations [Fig.~\ref{fig:structure}(d)], a magnetic moment of approximately 1 $\mu_B$ per unit cell is induced with the spin-up $p_z$ band fully filled and the spin-down one almost empty.
When spin-orbit coupling (SOC) is taken into account, the H-H Bi monolayer is a ferromagnetic insulator with the magnetization lying in the basal plane (x-y).
For the magnetization along +$\mathbf{z}$ axis, an SOC band gap of proximate 40 meV opens around the $K$ point, as shown in Fig.~\ref{fig:structure}(e). Using the Wannier interpolation method, we can calculate the band structure and the Berry curvature.  The Wannier band structure reproduces the FP one [Fig.~\ref{fig:structure}(e)] well. From the Berry curvature distribution in Fig.~\ref{fig:structure}(f), one can find an obvious dip around the valley $K$ and an imbalance between valley $K$ and $K'$, which indicates a nontrivial topological property of the bulk Bloch wave-functions. After integration of the Berry curvature throughout the whole Brillouin zone as well as around each individual valley, we obtain the Chern number~\cite{Chern} $C=-1$ as well as $C_K\simeq-1$ and $C_{K'}\simeq0$, demonstrating its nontrivial topological features of simultaneously possessing both QAH and QVH phases. This is further confirmed by the analysis of edge states, band evolution mechanism and spin texture, which will be discussed later.

\begin{figure*}
\includegraphics[width=7in]{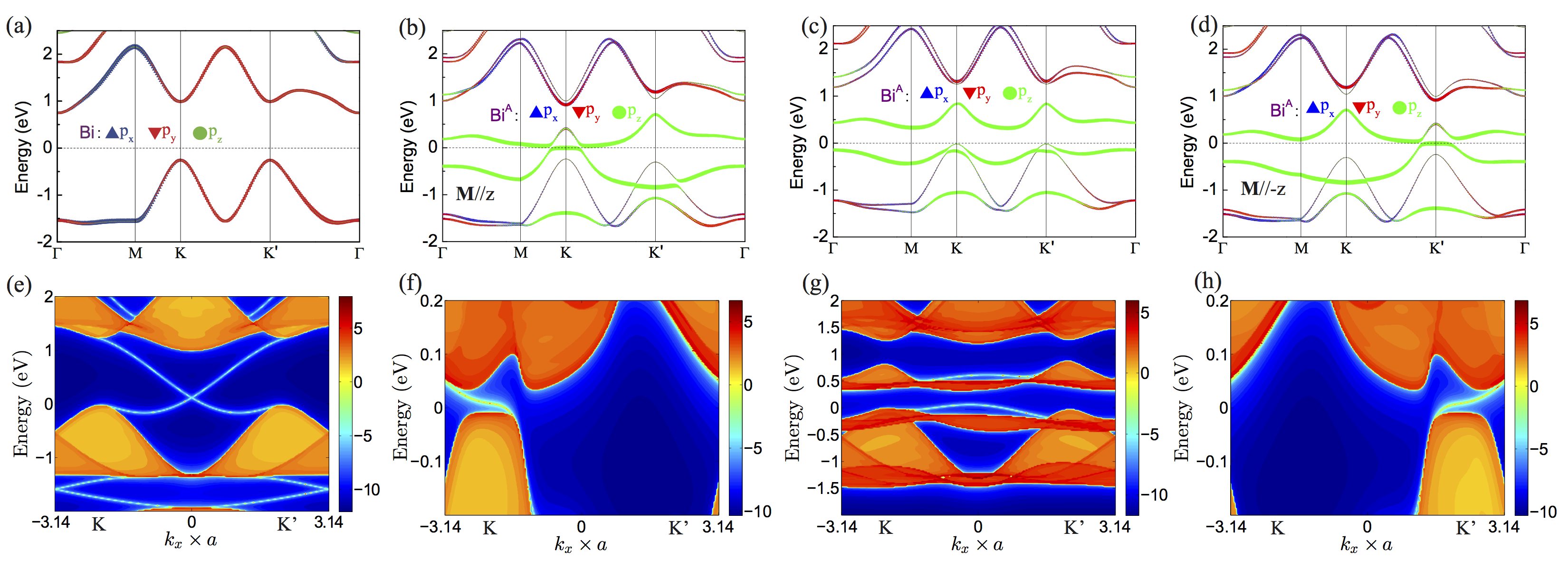}
\caption{(color online). (a) The projection band structures for the F-H Bi monolayer. The band edges mainly come from $p_x$ and $p_y$ orbitals from the Bi atoms of both sublattices.
(b)-(d) The projection band structures for the H-H Bi monolayer with the magnetization along the +$\mathbf{z}$ axis, in plane, and -$\mathbf{z}$ axis, respectively.  The bands around Fermi level mainly consist of the $p_z$ orbital from the dehydrogenated Bi atoms of the A sublattice. The energy spectrum for the semi-infinite zigzag monolayer in (e)-(h) correspond to the bulk spectrum in (a)-(d), respectively. There are two helical edges states in (e), indicating the QSH phase. (f)((h)) clearly presents the $K$ ($K'$) valley polarized chiral edge state with only a left (right) mover, hence is a VP-QAH1(2) phase. Nevertheless, (g) gives trivial edge states for ferromagnetic insulators.} \label{fig:edge_state}
\end{figure*}

In addition to the Chern number, the gapless edge mode inside the bulk energy gap provides a more intuitive picture to characterize the topological properties of the bulk. In Fig.~\ref{fig:edge_state}, we plot the band structures with their orbital projected character and the corresponding edge states of zigzag semi-infinite systems for four different cases. For the F-H Bi monolayer, the low-energy bands are dominated by the $p_x$ and $p_y$ orbitals of the Bi atoms as shown in Fig.~\ref{fig:edge_state}(a), and there are two helical edge states in the huge bulk gap of about 1 eV  [Fig.~\ref{fig:edge_state}(e)], which indicates a QSH phase.  For the H-H Bi monolayer, the two relevant flat bands mainly consist of the $p_z$ orbital from the dehydrogenated-site Bi atoms, which are located in the original large QSH gap of the F-H monolayer. Moreover, the orientation of the magnetization dramatically changes the relativistic band structures, especially the two relevant bands straddling the Fermi level,  as well as the size and the position of the gaps, as shown in Fig.~\ref{fig:edge_state}(b-d). Three typical insulating phases emerge for different orientation alignments of the magnetization. From the edge states of the zigzag semi-infinite systems and the direct calculations of Chern number [Fig.~\ref{fig:edge_state}(f-h)], we find the three typical insulating phases correspond to the VP-QAH1 ($C=-1$,$C_v=-1$), ferromagnetic insulator, and VP-QAH2 ($C=1$,$C_v=-1$) states, respectively.

\begin{figure*}
\includegraphics[width=7in]{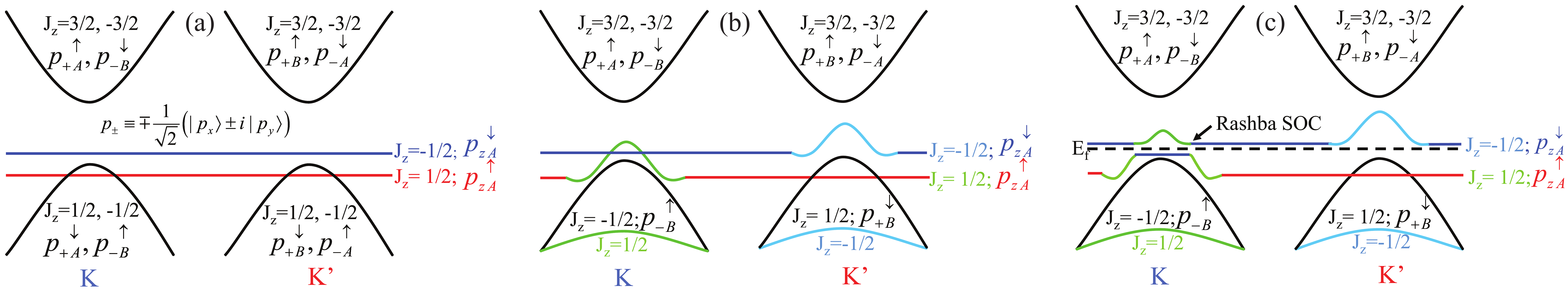}
\caption{(color online). (a)-(c) Schematic diagrams of the band evolution for the magnetization along the +$\mathbf{z}$ axis. The evolution stages are explained in the main text. } \label{fig:mechanism}
\end{figure*}

In order to obtain a physical picture of the nontrivial topological properties of VP-QAH, we investigate the band structure evolution and the spin texture as shown in Fig.~\ref{fig:mechanism}. Without loss of generality, we analyze the case of the magnetization aligned along the +$\mathbf{z}$ axis. The evolution is divided into three stages, as schematically plotted in Fig.~\ref{fig:mechanism}(a-c).
We start with the F-H Bi monolayer, whose low-energy band structure has a huge QSH band gap opened by on-site SOC~\cite{Liu2014}.
Dehydrogenation from the A sublattice leads to a flat band mainly consisting of the $p_z$ orbital from Bi atoms of the A-sublattice around the Fermi level in the otherwise huge QSH gap [Fig.~\ref{fig:structure}(c)].

\begin{figure}
\includegraphics[width=3.5in]{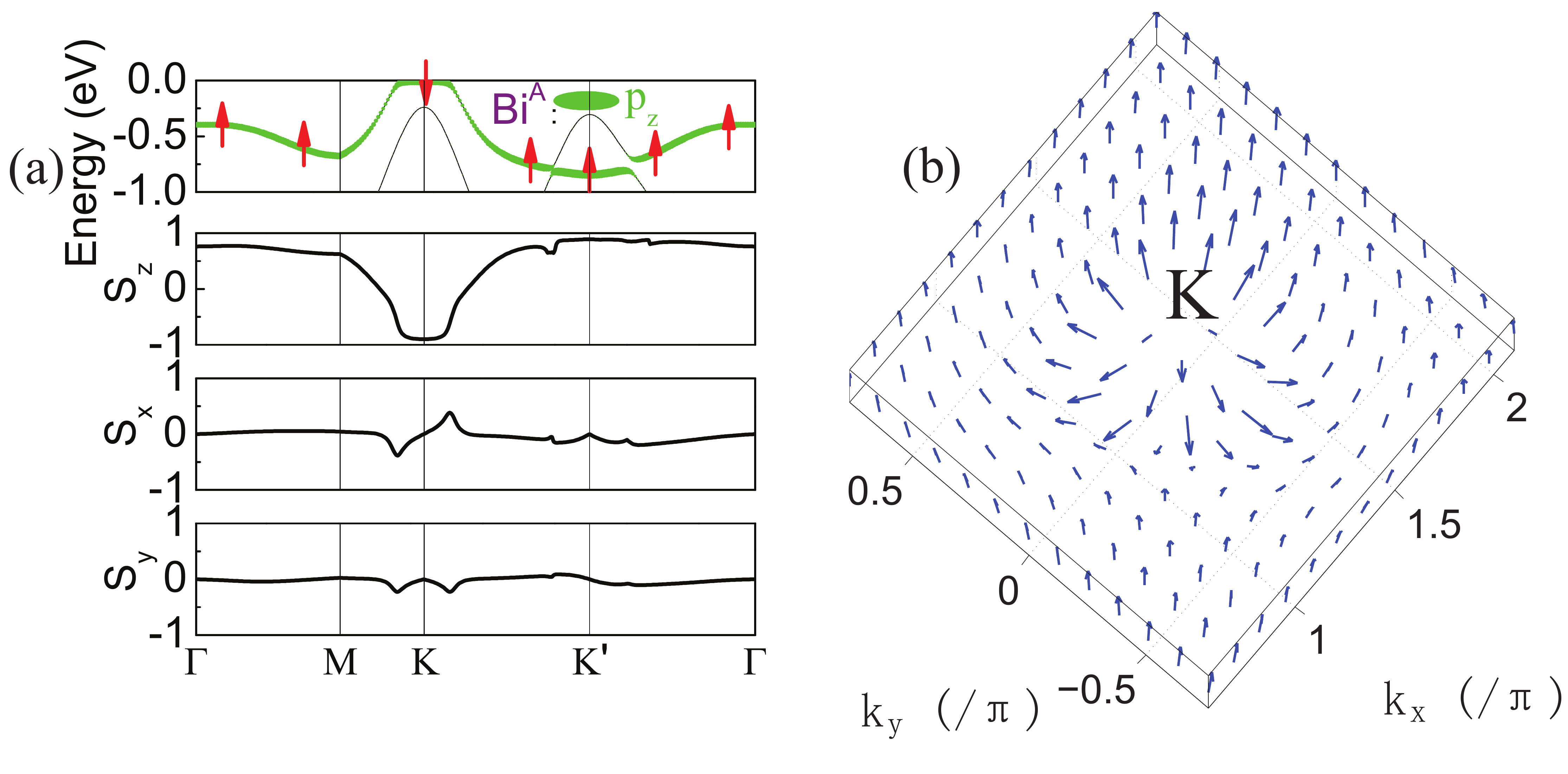}
\caption{(color online). (a)(b) 1D spin texture along the line of  high symmetry and 2D one around valley $K$ for the relevant $p_z$ orbital valence band from the dehydrogenated Bi atoms.  $S_x$, $S_y$, and $S_z$ are expectations of the three components of the spin operator for the $p_z$ orbital. The arrows represent the orientation of the spin, which is rotated by the intrinsic Rashba SOC and generates a nontrivial Skyrmion spin texture in the momentum space, i.e., around $K$ point, the spin points down, while far from $K$ point the spin points up, as plotted in (b) . The only one Skyrmion around $K$ results in a VP-QAH1 phase.} \label{fig:skyrmion}
\end{figure}

At the first stage of the band evolution, in Fig.~\ref{fig:mechanism}(a), the internal spontaneous magnetization splits the flat $p_{zA}$ bands into two bands, $p_{zA}^\uparrow$  and $p_{zA}^\downarrow$, where $p_{zA}$ means that the bands consist of the $p_z$ orbitals of the A-sublattice Bi atoms. Considering the electron filling, the Fermi level is located slightly lower than the $p_{zA}^\downarrow$ band. The intrinsic SOC plays a role in the second stage. From the pervious work~\cite{Liu2014}, it is known the bands with total angular momentum  $J_{z}^{\left\{ p_{x},p_{y}\right\} }=\pm1/2$  and $J_{z}^{\left\{ p_{x},p_{y}\right\}}=\pm3/2$ constitute the low-energy valence and conduction bands for QSH states in the F-H Bi monolayer respectively, as shown in \ref{fig:mechanism}(a). The superscript $\left\{ p_{x},p_{y}\right\}$ indicates the bands mainly come from the $p_{x},p_{y}$ orbitals. In the following, we will focus on the valence bands with $J_{z}^{\left\{ p_{x},p_{y}\right\}}=\pm1/2$, in view of that the total angular momentum conservation is required in the presence of SOC and the two relevant flat $p_z$ bands likewise own $J_{z}^{\left\{ p_{z}\right\}}=\pm1/2$.
Specifically, around the valley $K$, the valence bands consist of $p_{+A}^{\downarrow}$ ($J_z=1/2$) and $p_{-B}^{\uparrow}$ ($J_z=-1/2$), while around the valley $K'$, the subscripts sublattice index A and B are exchanged, where $p_{\pm}=\mp\left(p_{x}\pm ip_{y}\right)/\sqrt{2}$. As shown in Fig.~\ref{fig:mechanism}(b), the on-site (A-sublattice) SOC brings about the level repulsion
: $p_{+A}^{\downarrow}$ is pushed downward and $p_{zA}^\uparrow$ upward around valley $K$, while $p_{-A}^{\uparrow}$ is pushed downward and $p_{zA}^\downarrow$ upward around valley $K'$. At the last stage, in Fig.~\ref{fig:mechanism}(c), the intrinsic Rashba SOC breaks the band crossing and opens a gap around valley $K$, and results in a nontrivial state, i.e. VP-QAH state, which is validated from the spin texture for the relevant $p_z$ orbital valence band. As plotted in Fig.~\ref{fig:skyrmion}, the Skyrmion-type spin texture around valley $K$ can be mapped onto a whole spherical surface and thus gives rise to a nonzero winding number (Chern number) in the momentum space. Moreover, since there is only one Skyrmion around valley $K$,  rather than valley $K'$, the imbalance leads to a QVH effect. Consequently, a VP-QAH state emerges sharing novel properties of both QAH states and QVH states.

The effective minimal two-band Hamiltonian around two valleys $K$ and $K'$ is given to demonstrate the low-energy properties of the H-H Bi monolayer in the spin splitting basis $\left\{ |-M\rangle,|M\rangle\right\} $, which are linear combinations of $|p_{z}^{A},\uparrow\rangle$ and $|p_{z}^{A},\downarrow\rangle$ (for details see the Appendix A)
\begin{eqnarray}\label{heff}
\begin{split}
H_{\tau}^{eff}=& \left(\begin{array}{cc}
-2M & 0\\
0 & 0
\end{array}\right)+C_{k}\left(\begin{array}{cc}
1+\tau\cos\theta & -\tau\sin\theta e^{-i\phi}\\
-\tau\sin\theta e^{i\phi} & 1-\tau\cos\theta
\end{array}\right)\\
&+\frac{3}{2}\sin\left(\phi-\eta_{k}\right)akt_{R}\left(\begin{array}{cc}
-\sin\theta & f\\
f^{*} & \sin\theta
\end{array}\right),\\
\end{split}
\end{eqnarray}
where $C_{k}\equiv\lambda_{so}^{2}/\left(M+\sqrt{v_{f}^{2}k^{2}+\lambda_{so}^{2}}\right)$, and
$f\equiv e^{-i\phi}\left[i\cot\left(\phi-\eta_{k}\right)-\cos\theta\right]$.
$a$, $t_{R}$ and $2\lambda_{so}$ are the lattice constant, the strength of intrinsic Rashba SOC, and the intrinsic SOC strength, respectively.  $v_{f}$ is the Fermi velocity.  $M$, $\theta$ and $\phi$ are the respective strength, polar and azimuthal angles of
the magnetization.
$\eta_{k}$ is the angle between the vector $\mathbf{k}$ and the $\mathbf{x}$ axis. $\tau=\pm1$ labels two valleys $K$ and $K'$. The last term is the intrinsic Rashba SOC term.  By fitting the band structures using both FP and the above two-band model, these parameters can be determined as $v_{f}=1.1\times10^{6} m/s$, $\lambda_{so}=0.7 eV$, $M=0.32 eV$, and $t_{R}=0.02 eV$.

In the second stage, for the +$\mathbf{z}$ axial magnetization case, the crossing of the $p_{zA}^\uparrow$ and $p_{zA}^\downarrow$  bands around valley $K$ (see Fig.~\ref{fig:mechanism}(b)) is critical, which provides a pivot for the subsequent intrinsic Rashba SOC. Both of  these two constituents result in an inverted band gap as shown in Fig.~\ref{fig:mechanism}(c). Whether the two bands are crossing depends on the order of the eigenvalues in Eq.(1) when it is considered without the Rashba SOC term. It leads to the relations between $M$ and $\lambda_{so}$,
 $0<\frac{M}{\lambda_{so}}<\frac{\sqrt{5}-1}{2}$.
When the magnetization deviates the +$\mathbf{z}$ axial direction, the x and y components of the Zeeman term will also break the above crossing, resulting in a trivial phase. However, the nontrivial properties remain unchanged as long as Rashba SOC dominates the x and y Zeeman components. This requirement makes further restrictions on the ratio of $M$ and $\lambda_{so}$. For the case where the magnetization is along the negative $\mathbf{z}$ axis, the spin splitting is reversed, and band evolution with similar mechanism takes place around the $K'$ valley.

\begin{figure}
\includegraphics[width=3.5 in]{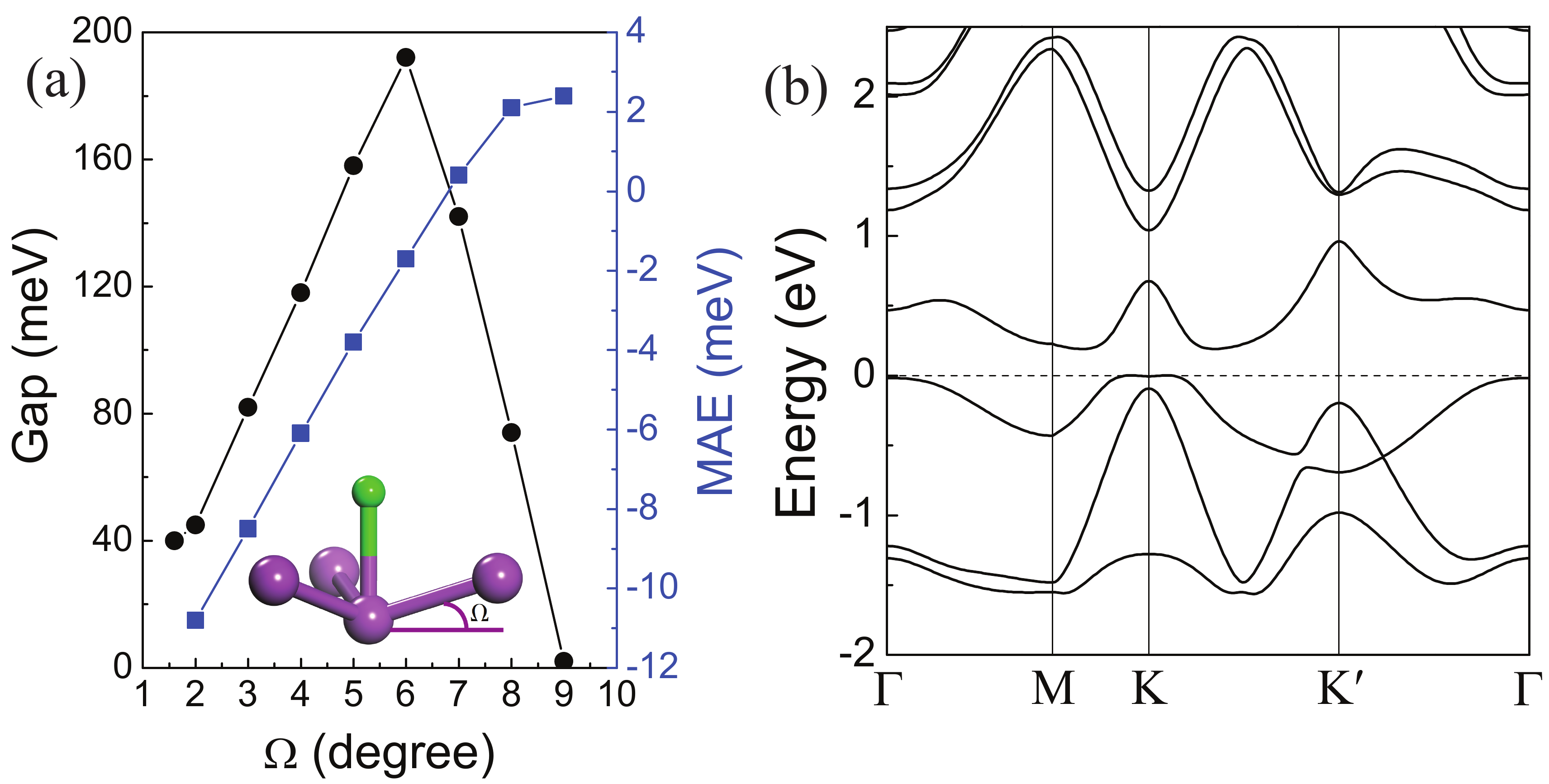}
\caption{(color online). The adiabatic evolution of the gap of VP-QAH and MAE. (a) The size of the gap (black line-dot) in the VP-QAH1 phase increases with the buckled angle firstly, as shown in the inset, and then decreases.  The MAE is defined as $MAE=E(\mathbf{M}\bot\mathbf{z})-E(\mathbf{M}\Vert\mathbf{z})$. Its value (blue line-square)  changes from negtive to positive, which means the easy magnetization axis orientating from in-plane to out of plane. (b) The band for the VP-QAH1 with the largest gap (0.19 eV), corresponding to the buckled angle $\Omega=6^{\circ}$ case. }\label{fig:GapBuckled}
\end{figure}

Moreover, we find that the buckled degree of the H-H Bi monolayer can  enlarge the gap of the VP-QAH phase significantly and change the magnetic anisotropy energy (MAE) dramatically, as shown in Fig.~\ref{fig:GapBuckled}(a). The largest gap of the VP-QAH1 phase closes to 0.2 eV with the buckled angle $\Omega=6^{\circ}$, whose band is plotted in Fig.~\ref{fig:GapBuckled}(b). Here the MAE is about 1 meV, hence the manipulation of orientation of magnetization is possible with an experimental accessible induced magnetic field (~10 T).  If we continue to increase the angle, the gap deceases. A insulator-metal phase transition takes place with the buckled angle $\Omega=9^{\circ}$ (see the Appendix B).

\begin{figure}
\includegraphics[width=3.2 in]{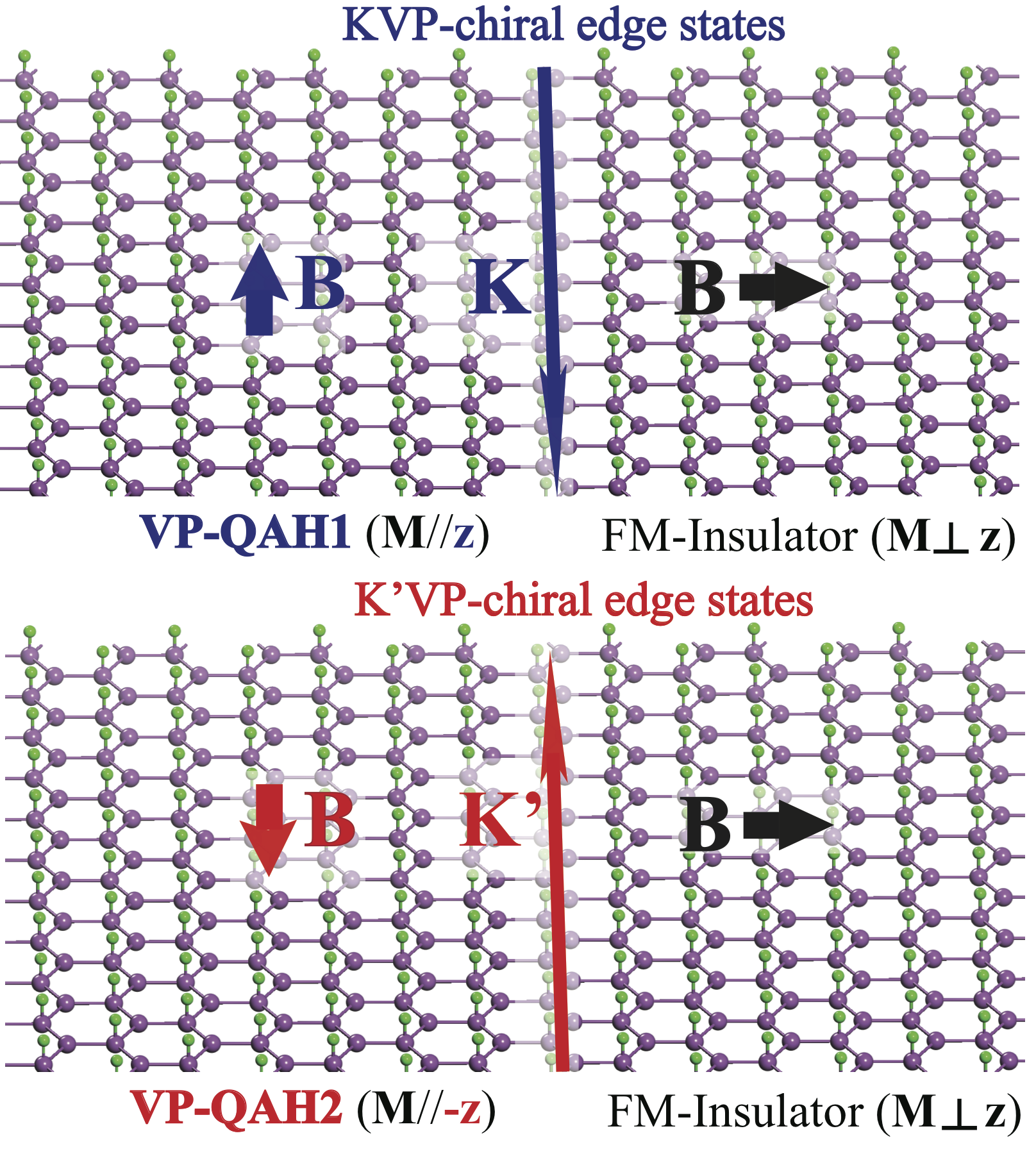}
\caption{(color online). A schematic diagram for chiral interconnects in a homogeneous junction made of an atomically thin H-H Bi monolayer, by tuning the orientations of magnetization. In the top (bottom) panel, along the borders of VP-QAH1 (VP-QAH2) insulators and FM insulators,  fully valley-polarized chiral edge states marked by valley $K$ ($K'$) index moves downwards (upwards)  with the external magnetic field in the left region pointing up (down). Such tunable valley-polarized chiral edge states can be used as dissipationless conducting wires for electronics and valleytronics.}\label{fig:Application}
\end{figure}

As one of the potential applications, it is possible to make a homogeneous junction by applying external magnetic fields with different orientations or by virtue of magnetic substrates at different regions of a H-H Bi monolayer film sample, as shown in Fig.~\ref{fig:Application}. Fully valley-polarized chiral edge states along these boundaries emerge and can be utilized as dissipationless conducting wires and chiral interconnects to lower the power consumption of devices in electronics and valleytronics. 

On the experimental side, Bi monolayers with a buckling honeycomb lattice and films have been manufactured via molecular beam epitaxy~\cite{Hirahara2011,Yang2012,Sabater2013}. On the other hand, chemical functionalization of such 2D materials is a powerful tool to create new materials with desirable features, such as modifying graphene into graphone (semihydrogenated graphene), graphane and fluorinated graphene using H and F, respectively~\cite{Zhou2009,Feng2012,Sofo2007,Elias2009,Robinson2010}. The buckled honeycomb geometry of the Bi monolayer makes it possible to saturate the chemical bonds of Bi atoms on only one sublattice. Therefore, it is a very promising that a H-H Bi honeycomb monolayer may be synthesized by chemical reaction in solvents or by exposure of a Bi monolayer or ultrathin film to atomic or molecular gases. An candidate substrate here could be layer materials such as transition metal dichalcogenides (e.g. MoS$_2$) and $\emph{h}$-BN, which stabilize and weakly interact with the H-H Bi honeycomb monolayer through van der Waals interaction and thus do not alter the band topology of the monolayer.

\section{CONCLUSION}

In summary, we find topological phase transitions in a H-H Bi honeycomb monolayer via tuning the orientation of the magnetization. Depending the orientation of the magnetization, there are four different phases, i.e. VP-QAH1, VP-QAH2, FM-Insulator, and FM-Metal, as shown in Fig.~\ref{fig:phase_diagram}. The mechanism for the nontrivial topological phase is given and a low-energy effective Hamiltonian is provided to capture the essential physics. Further, the low buckled geometry prominently increases the size of the gap by several times. Fully valley-polarized chiral edge states  can be utilized as dissipationless conducting wires and chiral interconnects for the lower power-consumption devices in electronics and valleytronics. These make the hydrogenated Bi honeycomb monolayers an ideal platform to investigate SOC relevant physics, novel topological states and the related phase transitions, and indicate great potential for the practical applications in a controllable manner.

\begin{acknowledgments}
This work was supported by the MOST Project of China (Nos. 2014CB920903, 2013CB921903, and 2011CBA00108), the National Natural Science Foundation of China (Grant Nos. 11404022, 11225418, and 11174337 ), the Specialized Research Fund for the Doctoral Program of Higher Education of China (Grant No. 20121101110046), and Excellent young scholars Research Fund of Beijing Institute of Technology (Grant No. 2014CX04028).
\end{acknowledgments}

\appendix

\section{The low-energy effective Hamiltonian}
From the first-principles (FP) calculations in the main text, the basis $\left\{ |p_{y}^{A}\rangle,|p_{x}^{A}\rangle,|p_{y}^{B}\rangle,|p_{x}^{B}\rangle,|p_{z}^{A}\rangle\right\} \otimes\left\{ \uparrow,\downarrow\right\}$ are relevant. The lattice Hamiltonian reads
\begin{eqnarray}\label{H}
\begin{split}
H& =\sum_{\langle i,j\rangle;\alpha,\beta=p_{x},p_{y}}t_{ij}^{\alpha\beta}c_{i\alpha}^{\dagger}c_{j\beta} \\
& +\sum_{i;\alpha,\beta=p_{x},p_{y};\sigma,\sigma^{'}=\uparrow,\downarrow}\lambda_{\sigma,\sigma^{'}}^{\alpha\beta}c_{i\alpha\sigma}^{\dagger}c_{i\beta\sigma^{'}}s_{\sigma,\sigma^{'}}^{z}  \\
&+\sum_{i\in A;\alpha=p_{z};\sigma,\sigma^{'}=\uparrow,\downarrow}c_{i\alpha\sigma}^{\dagger}c_{i\alpha\sigma^{'}}\left(\mathbf{s}\cdot\mathbf{M}\right) \\
& +\sum_{i\in A;\alpha\in\{p_{x},p_{y}\},\beta=p_{z};\sigma,\sigma^{'}=\uparrow,\downarrow}\lambda_{\sigma,\sigma^{'}}^{\alpha\beta}c_{i\alpha\sigma}^{\dagger}c_{i\beta\sigma^{'}}s_{\sigma,\sigma^{'}}^{z} \\
&+it_{R}\sum_{\langle\langle i,j\rangle\rangle\in A;\alpha=p_{z};\sigma,\sigma^{'}=\uparrow,\downarrow}c_{i\alpha\sigma}^{\dagger}c_{j\alpha\sigma^{'}}\left(\mathbf{s}\times\hat{\mathbf{d_{ij}}}^{0}\right)^{z} \\ &+t\sum_{\langle\langle i,j\rangle\rangle\in A;\alpha=p_{z}}c_{i\alpha}^{\dagger}c_{j\alpha} \\
& +\varepsilon_{p}\sum_{i;\alpha=p}c_{i\alpha}^{\dagger}c_{i\alpha}.\\
\end{split}
\end{eqnarray}
The first two terms represent the quantum spin Hall phase lattice Hamiltonian for the fully-hydrogenated (F-H) Bi honeycomb monolayer family~\cite{Liu2014}, which serves as the background for the following analysis of the half-hydrogenated (H-H) Bi monolayer. The third term is the Zeeman term for the $p_z$ orbital of Bi atoms of the dehydrogenated sites (A sublattice). The fourth term is the SOC between the $p_z$ and $p_x, p_y$ orbitals from A-site Bi atoms. The fifth term is the intrinsic Rashba SOC for the $p_z$ orbital from the dehydrogenated sites, as a result of the broken mirror symmetry. The sixth term is the hopping of the $p_z$ orbital from the dehydrogenated site, which is very small and thus can be ignored due to the next nearest neighbor (NNN) hopping. The last is the on-site energy term for the $p$ orbitals.

Firstly, the spontaneous magnetization $\mathbf{M}=M\left(\sin\theta \cos\phi,\sin\theta \sin\phi,\cos\theta\right)$ will lead to spin splitting with two eigenvalues -M and M, as shown in Fig. 4(a) in the main text, and their corresponding eigenstates
\begin{equation}\label{eigenstate}
\begin{split}
&|-M\rangle=\cos\frac{\theta}{2}|p_{z}^{A},\uparrow\rangle+\sin\frac{\theta}{2}e^{i\phi}|p_{z}^{A},\downarrow\rangle, \\
&|M\rangle=-\sin\frac{\theta}{2}e^{-i\phi}|p_{z}^{A},\uparrow\rangle+\cos\frac{\theta}{2}|p_{z}^{A},\downarrow\rangle, \\
\end{split}
\end{equation}
where $\theta$  and $\phi$  are the polar and azimuthal angles of the spontaneous magnetization $\mathbf{M}$.

Secondly, as stated above, for the QSH state, the SOC gap is huge (>1eV), and thus treated as the background in the following analysis. It's known that around the $K$ point, the dispersion is $E_{\pm}\left(k\right)=\pm\sqrt{v_{f}^{2}k^{2}+\lambda_{so}^{2}}$ , and the basis is as shown in Fig.~\ref{fig:mechanism}(a). The SOC mixes the orbital and spin with total angular momentum conserved $J_{z}=1/2$, which results in the level repulsion between $p_{zA}^{\uparrow} (|-M\rangle,|M\rangle)$  and $p_{+A}^{\downarrow}$, thus pushs the $p_{zA}^{\uparrow}  (|-M\rangle,|M\rangle)$  upward and the $p_{+A}^{\downarrow}$ downward, as shown in Fig.~\ref{fig:mechanism}(b). It is reasonable to choose the basis $\left\{ |-M\rangle,|M\rangle,p_{+A}^{\downarrow}\right\}$  as a low-energy subspace manifold. The corresponding Hamiltonian reads
\begin{equation}\label{3bands}
\begin{split}
&h_{K}=\\
&\left(\begin{array}{ccc}
-M+\varepsilon_{p} & 0 & \sqrt{2}\cos\frac{\theta}{2}\lambda_{so}\\
0 & M+\varepsilon_{p} & -\sqrt{2}\sin\frac{\theta}{2}e^{i\phi}\lambda_{so}\\
\sqrt{2}\cos\frac{\theta}{2}\lambda_{so} & -\sqrt{2}\sin\frac{\theta}{2}e^{-i\phi}\lambda_{so} & \varepsilon_{p}+E_{-}\left(k\right)
\end{array}\right),
\end{split}
\end{equation}
with $\varepsilon_{p}$  the on-site energy for the $p$ orbitals. Through the down-folding procedure~\cite{Winkler_spin-orbit_2003}, the minimal two-band low-energy Hamiltonian is obtained in the representation $\left\{ |-M\rangle,|M\rangle\right\}$
\begin{eqnarray}\label{2bands}
\begin{split}
h_{K}^{eff} &=\left(\begin{array}{cc}
-2M & 0\\
0 & 0
\end{array}\right) \\
&+\frac{\lambda_{so}^{2}}{M+\sqrt{v_{f}^{2}k^{2}+\lambda_{so}^{2}}}\left(\begin{array}{cc}
2\cos\frac{\theta}{2}^{2} & -\sin\theta e^{-i\phi}\\
-\sin\theta e^{i\phi} & 2\sin\frac{\theta}{2}^{2}
\end{array}\right). \\
\end{split}
\end{eqnarray}
The Fermi level is taken as $\varepsilon_{p}+M$  during the above derivation.

\begin{figure}
\includegraphics[width=3.5 in]{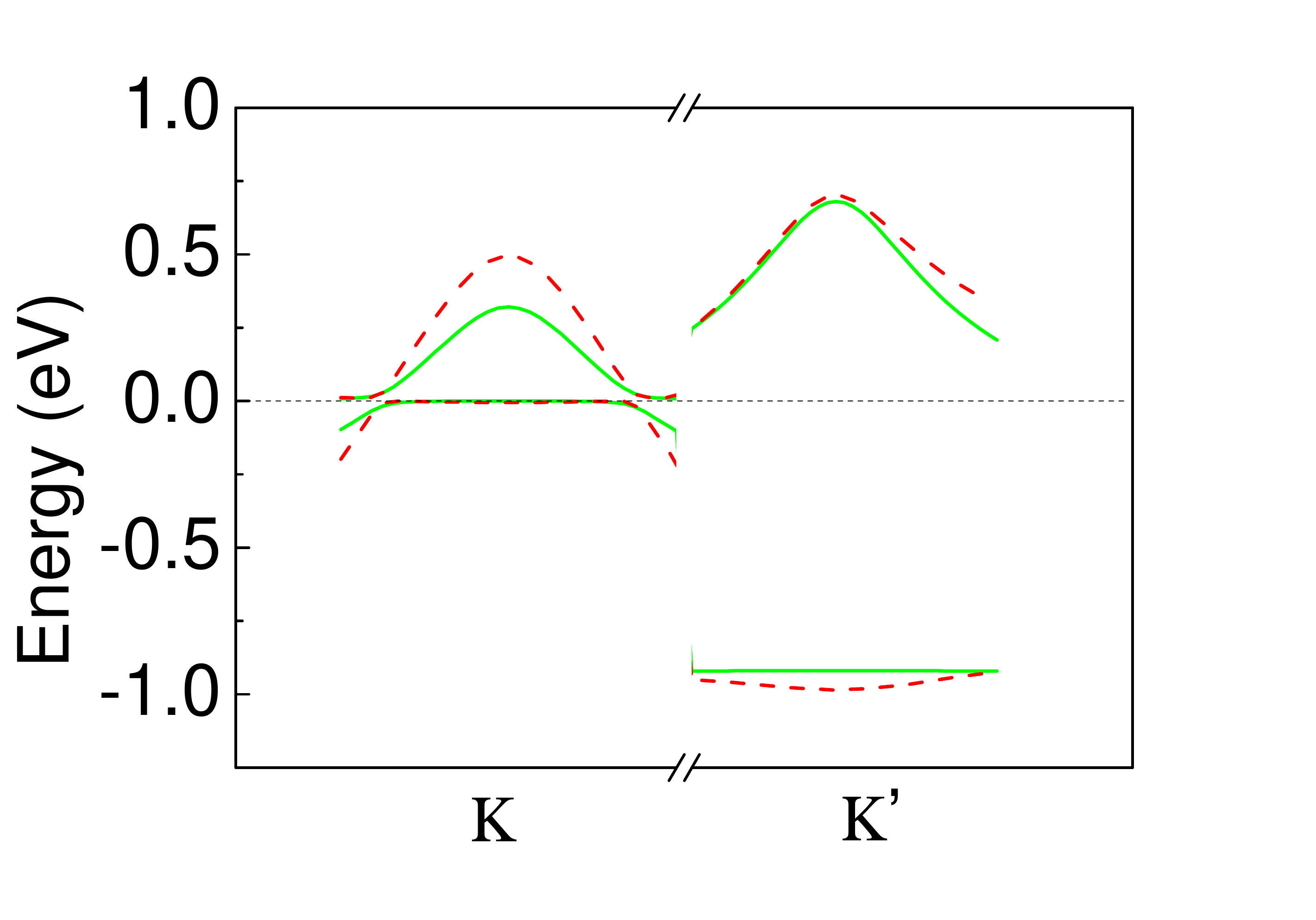}
\caption{(color online). The band structures around two valleys $K$ and $K'$ for the H-H Bi monolayer from FP calculation and the two-band low-energy effective Hamiltonian  with the magnetization along +$\mathbf{z}$ axis. The dashed red curve is the FP result. The solid green curve represents the TB model's result. The Fermi level is set to zero.}\label{fig:TBvsFP}
\end{figure}

At the last stage, as shown in  Fig.~\ref{fig:mechanism}(c) in the main text, the intrinsic Rashba SOC opens a band gap around valley $K$, and results in the nontrivial state. After performing a Fourier transformation, the Rashba SOC is written in the basis $\left\{ |-M\rangle,|M\rangle\right\}$
\begin{eqnarray}\label{Rashba}
\begin{split}
h_{K}^{Rashba}=& \frac{3}{2}\sin\left(\phi-\eta_{k}\right)akt_{R} \times \\
& \left(\begin{array}{cc}
-\sin\theta & f\left(\theta,\phi,\eta_{k}\right)\\
f\left(\theta,\phi,\eta_{k}\right)^{*} & \sin\theta
\end{array}\right),\\
\end{split}
\end{eqnarray}
with $f\left(\theta,\phi,\eta_{k}\right)\equiv e^{-i\phi}\left[i\cot\left(\phi-\eta_{k}\right)-\cos\theta\right]$.
$a$ and $t_{R}$ are the lattice constant and the strength of the Rashba SOC, respectively. $\eta_{k}$ is the angle between the vector $\mathbf{k}$ and the $\mathbf{x}$ axis.

Consequently, the total minimal two-band model around $K$ point reads
\begin{equation}\label{HK2bands}
H_{K}^{eff}=h_{K}^{eff}+h_{K}^{Rashba}.
\end{equation}

Following a similar three-step procedure for the minimal two-band model around $K'$ point, as shown in Fig.~\ref{fig:mechanism}(a-c)  in the main text, the total minimal two-band model in the representation $\left\{ |-M\rangle,|M\rangle\right\}$  around valley $K'$ is found to be
\begin{equation}\label{HK'2bands}
H_{K'}^{eff}=h_{K'}^{eff}+h_{K}^{Rashba},
\end{equation}
with
\begin{eqnarray}\label{hK'}
\begin{split}
h_{K'}^{eff} &=\left(\begin{array}{cc}
-2M & 0\\
0 & 0
\end{array}\right) \\
&+\frac{\lambda_{so}^{2}}{M+\sqrt{v_{f}^{2}k^{2}+\lambda_{so}^{2}}}\left(\begin{array}{cc}
2\sin\frac{\theta}{2}^{2} & \sin\theta e^{-i\phi}\\
\sin\theta e^{i\phi} & 2\cos\frac{\theta}{2}^{2}
\end{array}\right).\\
\end{split}
\end{eqnarray}
By fitting the band structures between FP and the above low-energy two-band model around the two valleys $K$ and $K'$, the above parameters are determined with $v_{f}=1.1\times10^{6} m/s$, $\lambda_{so}=0.7 eV$, $M=0.32 eV$, and $t_{R}=0.02 eV$, as shown in Fig.~\ref{fig:TBvsFP}.

\section{The adiabatic evolution of the gap of the VP-QAH phases with the buckled angle}

\begin{figure*}[!]
\centering
\includegraphics[width=7 in]{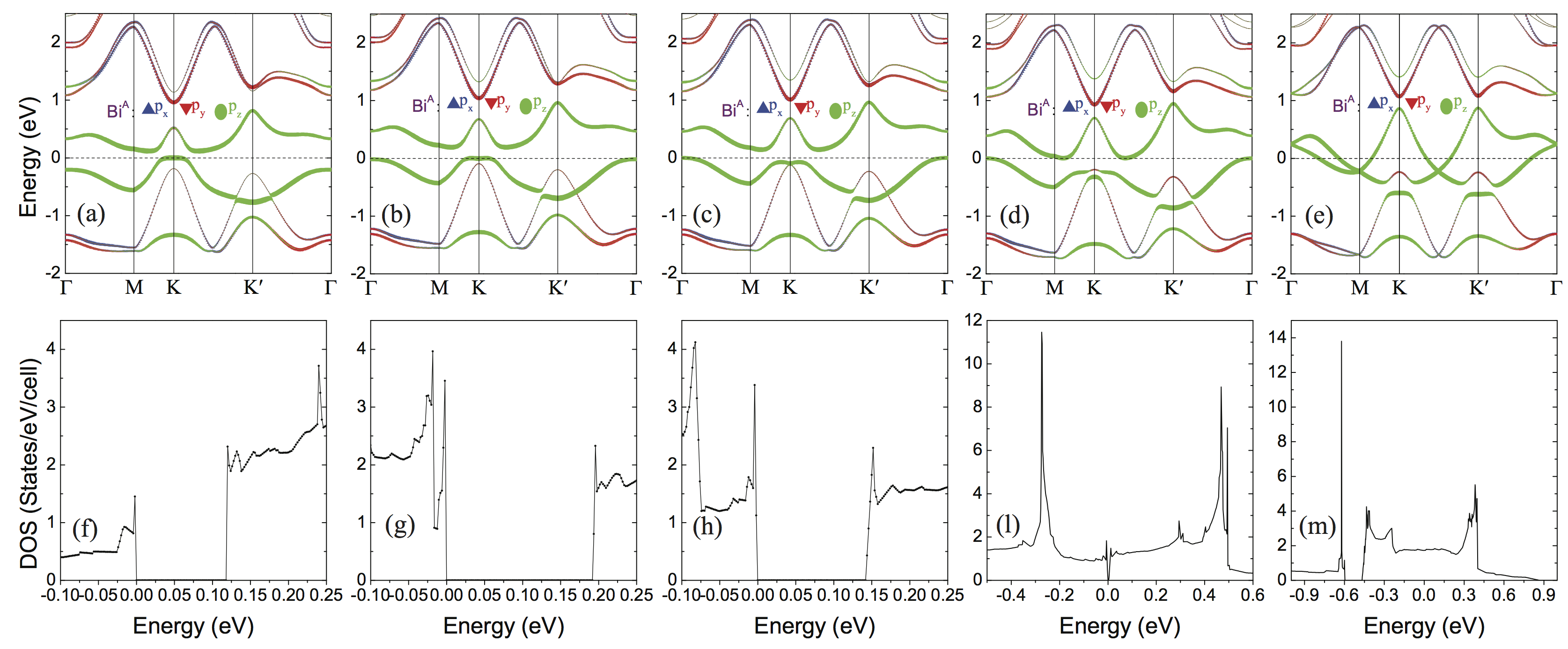}
\caption{(color online). (a)-(e) The projection band structures for the H-H Bi monolayer with buckled angles $\Omega=4^{\circ}, 6^{\circ}, 7^{\circ}, 9^{\circ}, 10^{\circ}$. The color of the symbols labels the different atomic orbitals, and their size is proportional to the weight of the band eigenfunctions on these atomic orbitals.  (f)-(m) The respective density of states correspond to (a)-(e).}\label{fig:Angle}
\end{figure*}

We investigate the adiabatic evolution of the gap of the VP-QAH from the quasi-planar honeycomb geometry to the low-buckled geometry with keeping the Bi-Bi bond length constant, as shown in Fig.~\ref{fig:GapBuckled} in the main text. It is surprising that the slightly buckled geometry remarkably magnifies the size of the gap (even reach 0.19 eV), which make the giant-gap H-H Bi monolayer an ideal platform to realize the exotic VP-QAH phases and fabricate new quantum devices operating at room temperature. The projected band structures and density of states for the four typical points ($\Omega=4^{\circ}, 6^{\circ}, 7^{\circ}, 9^{\circ}$) during the gap evolution as well as the $\Omega=10^{\circ}$ case are plotted in Fig.~\ref{fig:Angle}.

\bibliography{}

\end{document}